\documentclass[useAMS,usenatbib]{mn2e}

\usepackage{graphicx}

\title[Silicate absorption in galaxy nuclei]{Silicate absorption in heavily obscured galaxy nuclei}
\author[P.F. Roche et al.]{Patrick F. Roche$^{1}$\thanks{E-mail:
p.roche@physics.ox.ac.uk },  Christopher Packham$^{2}$, David K. Aitken$^3$,   Rachel E. Mason$^{4}$, 
 \\
$^{1}$Astrophysics, University of Oxford, Dept of Physics, DWB, Keble Road Oxford OX1 3RH\\
$^{2}$Department of Astronomy, University of Florida, 211 Bryant Space Science Center, PO Box 112055, Gainesville, Fl 32611-2055, USA. \\
$^{3}$Dept.Physical Sciences,  University of Hertfordshire, College Lane, Hatfield, Herts, AL10 9AB. \\
$^{4}$Gemini Observatory, 670 N. A'ohuku Place, Hilo, Hawaii 96720, USA.. \\
}
\begin{document}

\date{Accepted 2006 October 17 . Received 2006  October 17; in original form 2006 September 12}

\pagerange{\pageref{firstpage}--\pageref{lastpage}} \pubyear{2006}

\maketitle

\label{firstpage}

\begin{abstract}

Spectroscopy at 8-13$\umu$m with T-ReCS on Gemini-S is presented for 3 galaxies with substantial silicate absorption features, NGC 3094, NGC 7172 and NGC 5506.  In the galaxies with the deepest absorption bands, the silicate profile towards the nuclei is well represented by the emissivity function derived from the circumstellar emission from the red supergiant,  $\mu$ Cephei which is also representative of the mid-infrared absorption in the diffuse interstellar medium in the Galaxy.  There is spectral structure near 11.2$\umu$m in NGC 3094 which may be due to a component of crystalline silicates.  In NGC 5506, the depth of the silicate absorption increases from north to south across the nucleus, suggestive of a dusty structure  on scales of 10s of parsecs.  We discuss the profile of the silicate absorption band towards galaxy nuclei and the relationship between the 9.7~$\mu$m silicate and 3.4~$\mu$m hydrocarbon absorption bands.

\end{abstract}


\section{Introduction}

Observations of active galaxies have shown  that there are very large columns of material along the line of sight to many  nuclei.  The standard picture is that Seyferts of type 2 suffer significantly greater obscuration than those of type 1, much of which probably arises in dusty circumnuclear structures with differing inclinations to the line of sight (eg. Antonucci 1993).   These columns can be investigated via the cut-offs at  X-ray wavelengths, the extinction in the visible and near-infrared, and dust absorption bands in the near and mid-infrared, all of which probe to different depths.  
X-ray observations have shown that at least 50\% of nearby Seyfert 2 galaxies are obscured by  column densities greater than 10$^{24}$ cm$^{-2}$ (Risaliti, Maiolini, \& Salvati 1999) and that this material is concentrated into relatively small volumes ($<$ a few parsecs).   Mid-infrared observations have demonstrated that substantial  silicate absorption depths at 9.7~$\umu$m  are relatively common towards Seyfert 2 galaxies, but rare in Seyfert of type 1 (Aitken \& Roche 1985, Roche et al 1991), again suggesting large obscuring columns towards the underlying warm emitting dust. 

Dust around AGN is likely to be affected by the hard photon flux from the active nucleus;  for example Aitken \& Roche (1985) have argued that the absence of PAH emission bands in active nuclei can be explained by destruction of the small grains that carry the bands.  On the other hand, the shape of the silicate absorption feature towards the most heavily extinguished galaxy known, NGC 4418 (Roche et al 1986), is similar to that produced by the diffuse interstellar medium in our Milky Way galaxy (Roche \& Aitken 1984, 1985, Chiar \& Tielens 2006).  A weak absorption band at 3.4$\umu$m, similar to that seen in the Galactic diffuse ISM, has been detected towards a number of nuclei (Imanishi 2000, 2003, Mason et al 2004), again suggesting that the dust towards those objects is similar to that in the Galaxy.  Observations with Spitzer have recently detected absorption features from crystalline silicates towards several deeply-embedded galaxies and additional absorption bands from hydrocarbons at 6.85 and 7.25~$\umu$m and a number of molecular species (Spoon et al 2006).

Here we present mid-infrared spectroscopy of NGC 3094, NGC 7172 and NGC 5506, three galaxies with significant silicate absorption features, and investigate the nature of the silicate absorption towards active nuclei at subarcsecond resolution.

\begin{figure*}
   \centering
   \resizebox{\hsize}{!}{\includegraphics[bb=140 280 460 600, clip=true, angle=90]{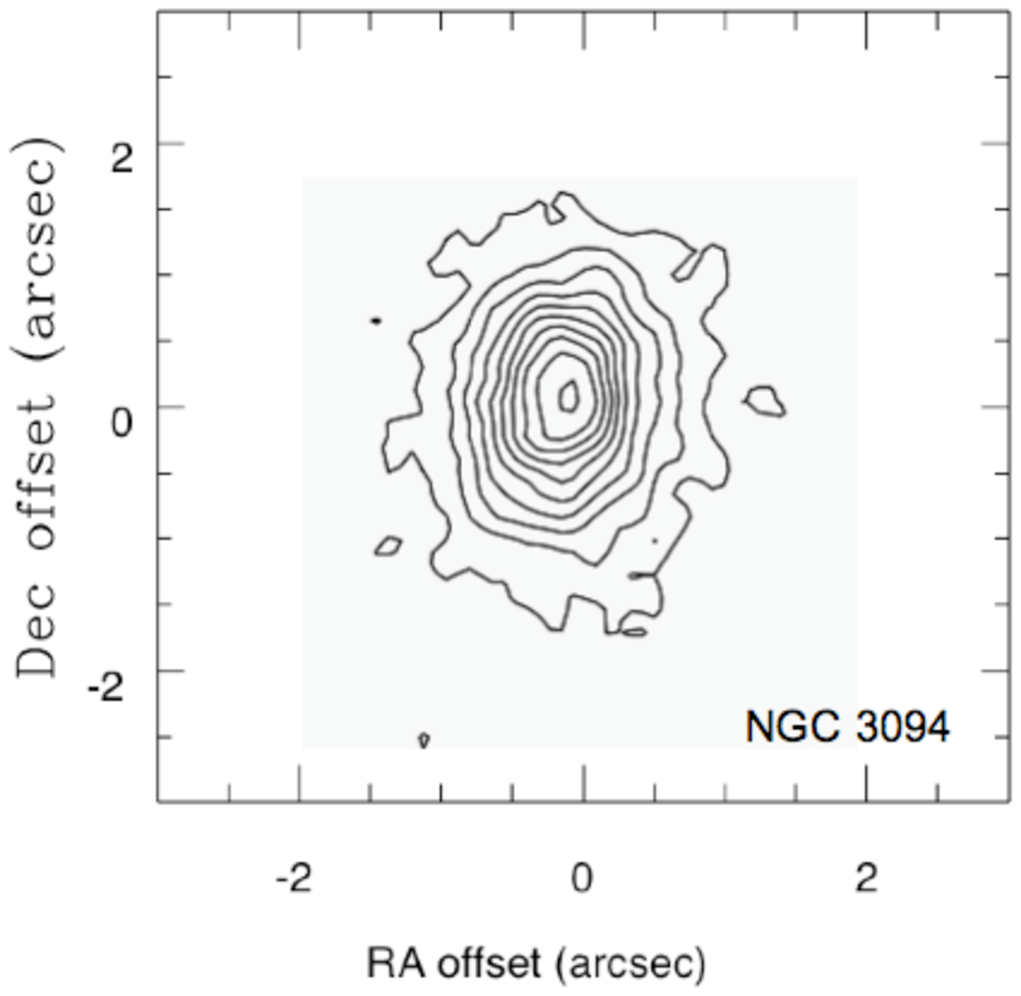}
   \includegraphics[bb=140 280 460 600, clip=true, angle=90]{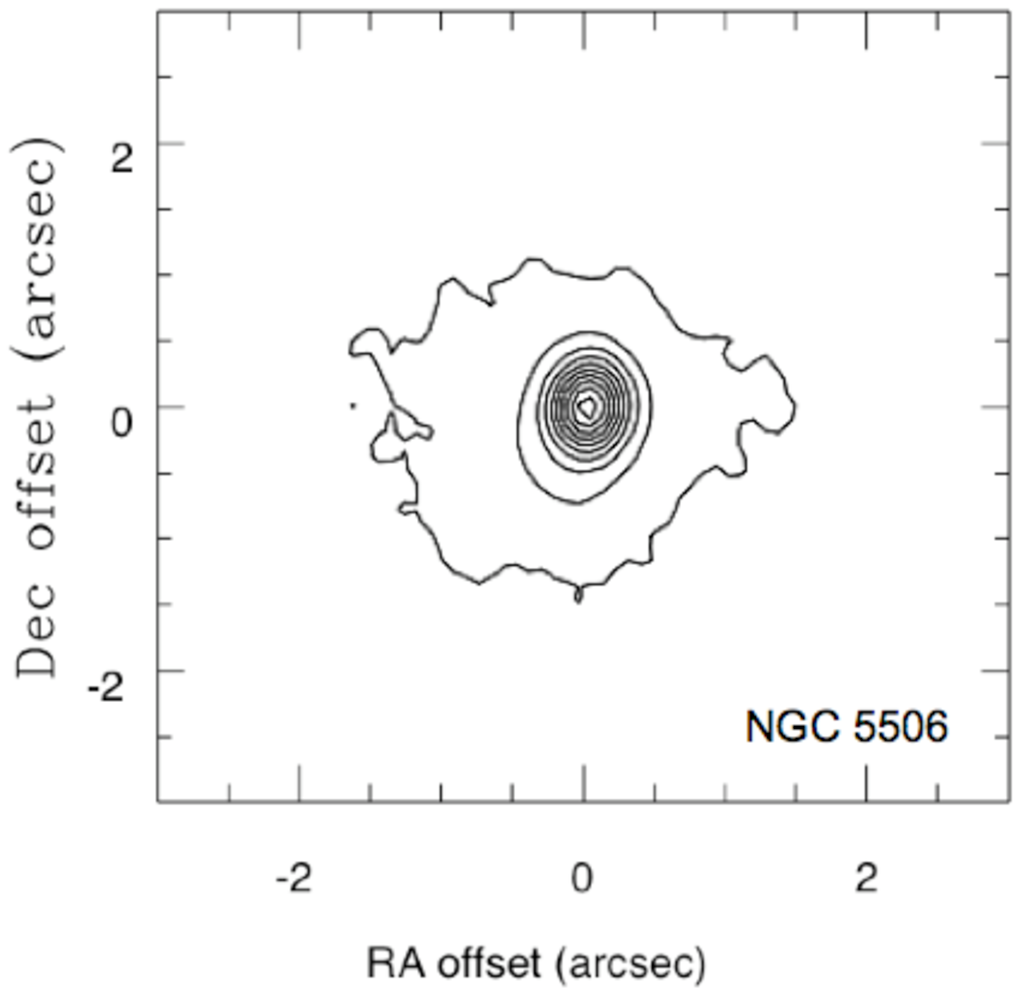}
   \includegraphics[bb=140 280 460 600, clip=true, angle=90]{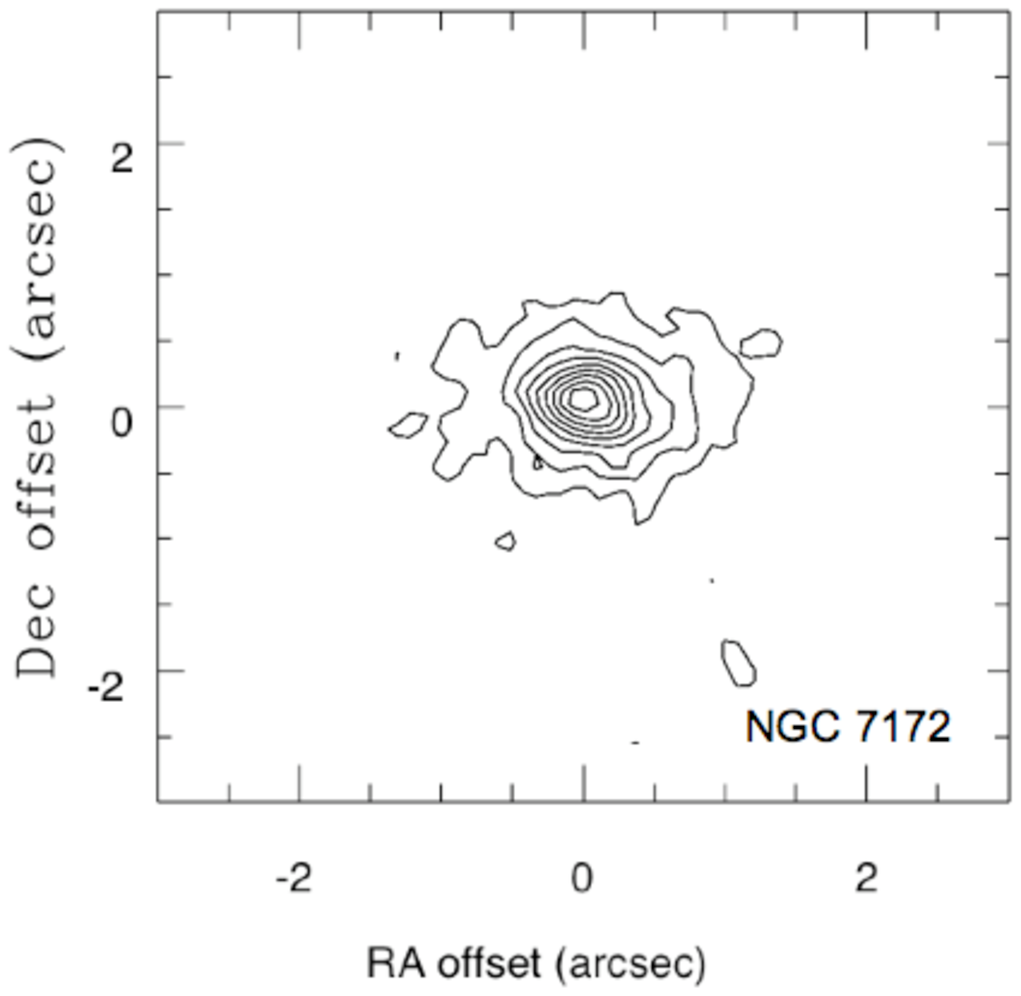}
}
     \caption{Acquisition images of NGC 3094, NGC 5506 and NGC 7172 at 12~$\mu$m.  The contours are linear, but the flux scale is arbitrary and positions are relative to the peak of the mid-IR flux.   Sky conditions were unstable during observations of NGC3094, and hence the small scale structure of this source my not be dependable.
     }
	\label{Acqimages}     
	\end{figure*}

\section[]{Observations}

Long slit  spectra between 8 and 13~$\umu$m were obtained at the 8-m Gemini South telescope in clearing skies with the facility mid-infrared imager/spectrometer, T-ReCS (Telesco et al 1998), in May 2004 under programme GS-2004A-C-2.  The seeing was variable at the time of observations, so the resolution achieved in the acquisition images is not well defined, particularly for NGC 3094 which was observed as the skies cleared and conditions were changing rapidly.   Nontheless, we show these short exposures in Figure 1.  We do not have accurate astrometric reference positions for the T-ReCS data  and so adopt the bright core as the central reference position.  The bright nuclei were centred in the slit before taking the spectroscopic observations.   

Spectra were obtained with standard chop and nod techniques (chop throw 15 arcsec) after centering the compact nucleus in the 0.36 arcsec wide slit.  The instrument was configured with the low resolution (11 line/mm) grating giving a dispersion of 0.0223 $\umu$m/pixel and a spectral resolution of 0.08$\umu$m. T-ReCS has a detector scale of 0.089 arcsec per pixel which provides a 25 arcsec long slit in the spatial direction and coverage of the full N photometric band, limited by the N filter bandpass, in the dispersion direction.  
Series of A B A B nod positions were accumulated, with total on-source integrations of 10, 10 and 5 minutes on source for NGC 3094, NGC 5506 and NGC 7172 respectively.  The chop direction was perpendicular to the slit, and so only the signal from the guided beam position was accumulated.

Correction for telluric atmospheric absorption is with respect to $\alpha$ Cen whilst  wavelength calibration is with respect to the planetary nebula NGC 6302, which has a rich emission line spectrum (see Casassus, Roche \& Barlow  (2000) for the spectrum and accurate wavelengths).

\begin{figure*}
   \centering
   \resizebox{\hsize}{!}{\includegraphics[clip=true]{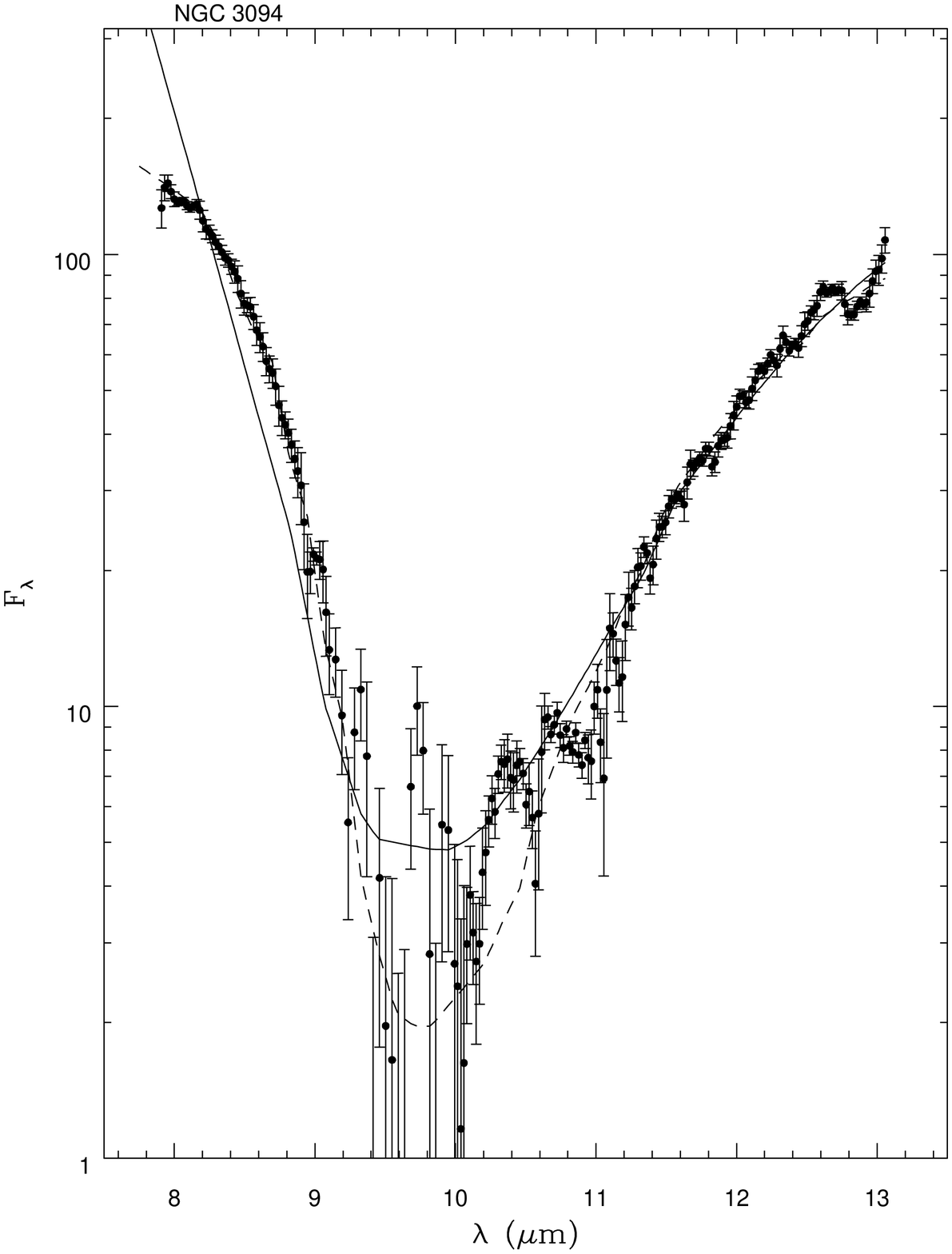} 
   \includegraphics[clip=true]{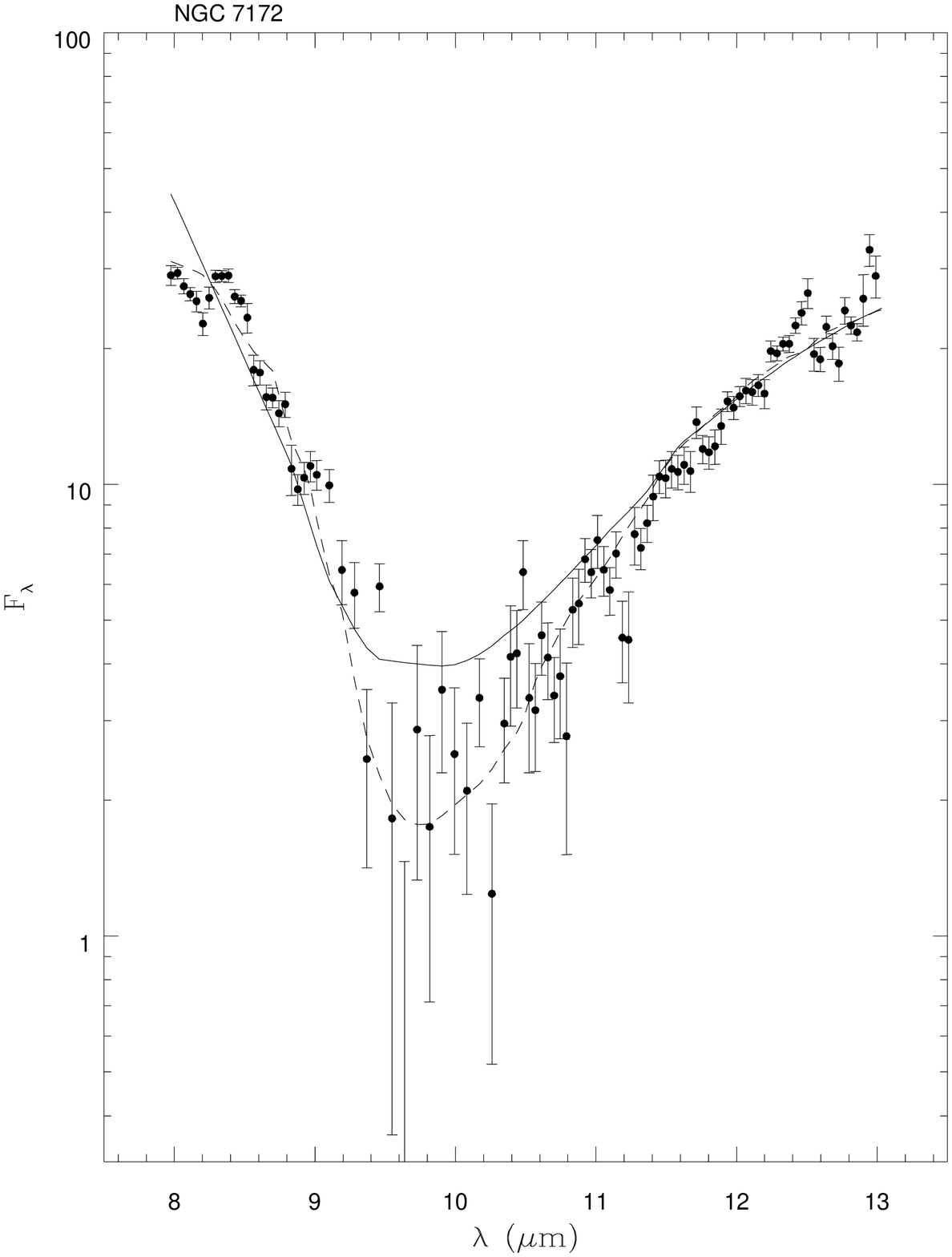}
   }
   \vspace*{-10mm}
     \caption{The 8-13~$\umu$m spectra of NGC 3094 (left) and NGC 7172 (right).  The solid lines represent fits to the spectra with the Trapezium silicate curve while the dashed lines use that obtained from the circumstellar shell of $\umu$ Cep.  Flux is in  units of 10$^{-20}$ W cm$^{-2} \umu$m$^{-1}$.  The error bars plotted are estimated from the scatter of the points in narrow wavelength bins. the  spectrum of NGC 7172 has been smoothed to improve the S/N ratio per point.  }
        \label{gal1}
    \end{figure*}

\begin{figure*}
   \centering
   \resizebox{\hsize}{!}{\includegraphics[clip=true]{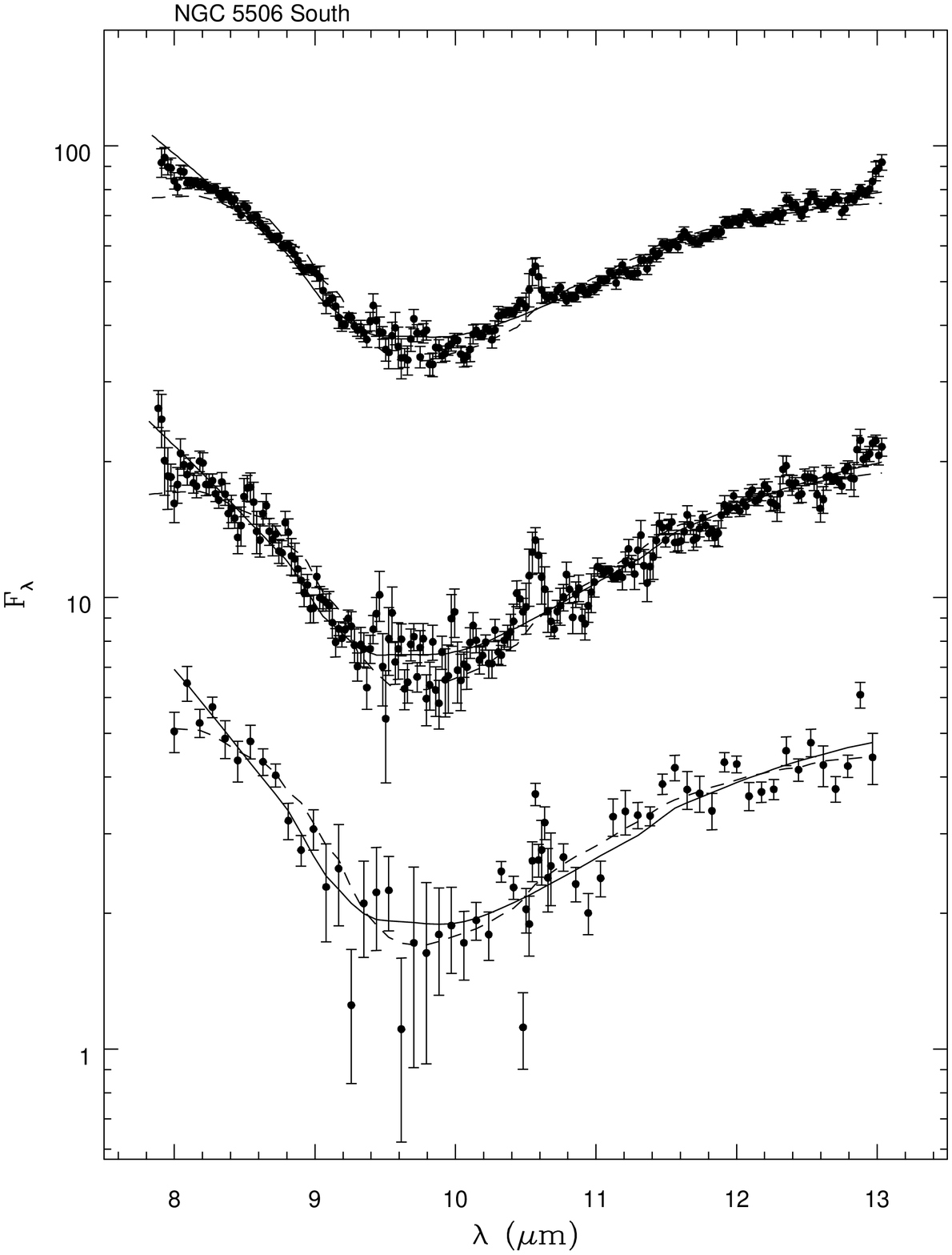}
 \includegraphics[clip=true]{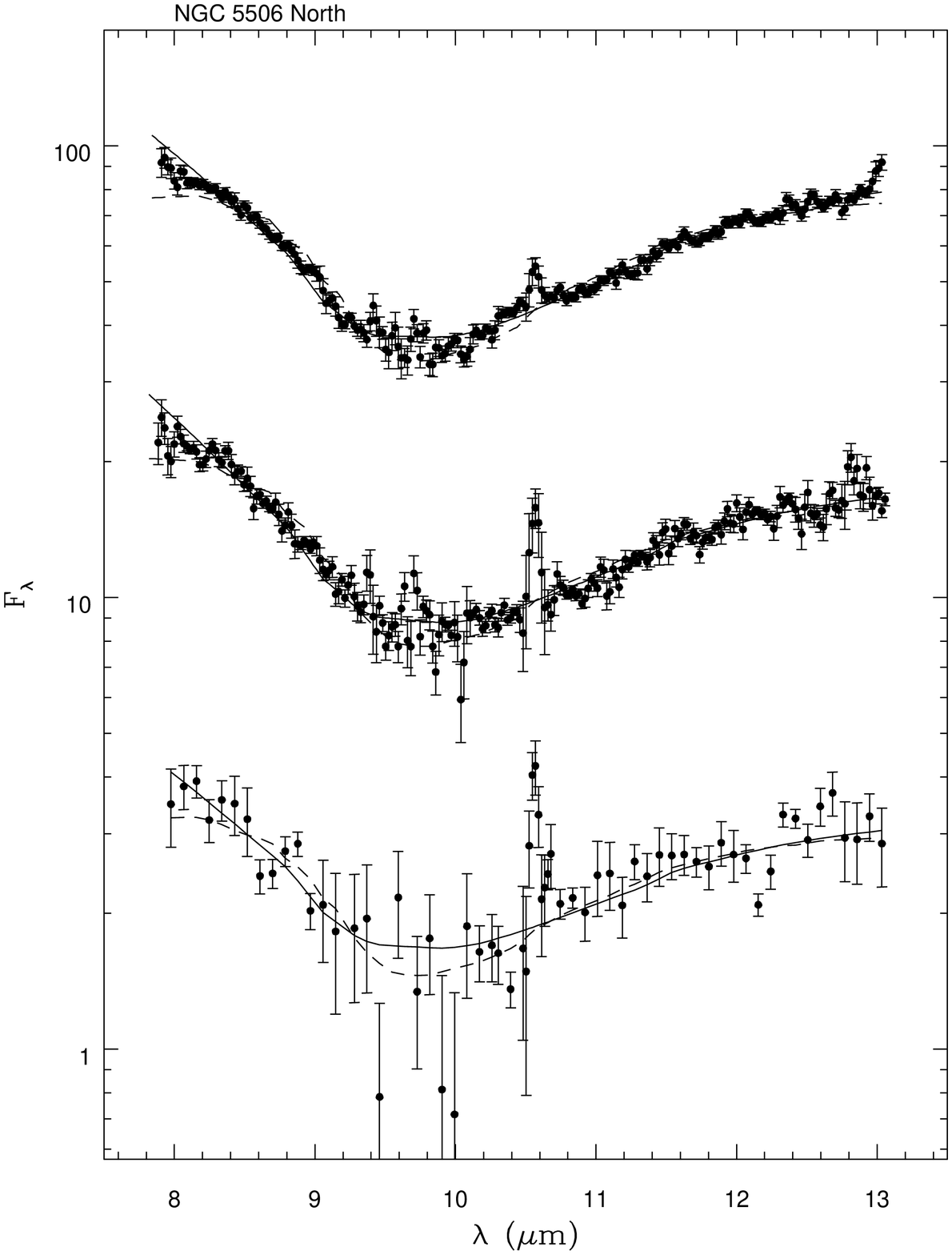}
}
     \caption{The spectra of NGC 5506. Left: the spectrum of the peak (top) and at positions 0.3 and 0.6 arcsec south. Right the spectrum at the peak and at positions 0.3 and 0.6 arcsec north of the centre.  Flux is in units of 10$^{-20}$Wcm$^{-2} \umu$m$^{-1}$ and the dashed and solid lines represent fits as described for figure 2.  The spectra at 0.6 arcsec S and N have been smoothed to improve the S/N ratio per point, except in the vicinity of the [S IV] emission line.  
}  
        \label{n5506}
    \end{figure*}

The spectra were straightened, registered in the dispersion direction with reference to the sharp structure in the absorption band due to atmospheric ozone at 9.6 microns, and the dispersion calibration from NGC 6302 was applied. The spectra were corrected for atmospheric absorption by straightening and growing the alpha Cen spectrum along the slit, dividing it into the galaxy spectra and multiplying by a blackbody at a temperature of 5770~K (Cohen et al 1996).   We extracted and smoothed the residual flux  in regions $\pm$5-10 arcsec along the slit from the spectra of the nuclei; this is at a very low level and presumably arises from beam imbalances between the chopped and nodded beams. The deep silicate absorption bands in NGC 3094 and NGC 7172 give rise to very low flux levels in the 9 to 10~$\umu$m region, and the shape of the minimum can be distorted by this residual background. We therefore subtracted the residual signal from the galaxy spectra. This only has a significant effect on the flux between 9.2 and 10$\umu$m, but does appear to improve the match to the expected silicate profiles.  

The cores of these galaxies are fairly compact and the slit losses for NGC 5506 and NGC 7172 are probably similar to those for the  reference star. Nonetheless,  the calibration uncertainty is likely to be significant and we estimate $\sim$20\%.  
%Details of the observations are given in Table 1. 

%\begin{center}
%Table 1.  Details of Observations

%\hspace*{3 mm}

%\begin{tabular} {l r l l}
%		 & [SIV] & [NeII] & 11.3$\umu$m \\[2 mm]
%NGC 3094 & 60$^\circ$ 	& 10$\pm$ 3 & \\
%NGC 5506  &  0$^\circ$ 	& 4.7$\pm$1 & 4.1$\pm$ 3 \\
%NGC 7172  &  100 $^\circ$  & 13 & 96  \\

%\end{tabular}
%\end{center}
%\noindent 
%Note: \\
%Intensities are summed along the inner 5 arcsec of the slits, but omitting the central 0.6 arcsec, and are in units of 10$^{-20}~$Wcm$^{-2}$ \\[5mm]

In order to search for spectral variations along the slit, spectra were extracted in a number of different spatial bins.  For NGC 3094 and NGC 7172, the extracted spectra off the nuclei were of low signal to noise, and no significant differences from the on-nuclei spectra  were detected.   NGC 5506 is significantly brighter, and its spectra do show spatial structure.   Spectra were extracted by summing detector rows together, and are shown in figure 3 at the position of the central peak, at $\pm$0.3 arcsec and at $\pm$0.6 arcsec   The noise in the spectra increases between 9 and 10~$\umu$m because of  low atmospheric transmission due to absorption by ozone in the upper atmosphere, while the fluxes at these wavelengths are decreased by silicate absorption in the source.  These effects become increasingly apparent in the spatial positions furthest from the nucleus.

\section{Targets}

The galaxies observed all show significant silicate absorption bands in published mid-infrared spectra with no evidence of the narrow emission bands attributed to emission by large polycyclic aromatic hydrocarbon molecules (PAHs)  that trace nuclear star-forming regions (Roche et al 1991),   and show other evidence of substantial obscuration.  They are described briefly here.

\subsection{NGC 3094}

NGC 3094 is a nearby (z$\sim$0.008) SBa spiral galaxy hosting a Seyfert 2 nucleus. and inclined by $\sim$49$^\circ$ to the line of sight.   Near-infrared images show a bright bar with a compact nucleus extended over $\sim$1.7 arcsec (Zenner \& Lenzen 1993) while the radio emission is fairly diffuse over ~15 arcsec  (Condon et al 1990).  The T-ReCS acquisition image indicates that the mid-infrared source may be extended by about 1 arcsec at a  position angle of $\sim170^\circ$, but the variable seeing conditions make this rather uncertain. However,  the spatial profile within the slit, which was at position angle 0$^\circ$ for the spectroscopic observations also suggests a source extended on a scale of $\sim$1 arcsec     Despite the narrow slit, the T-ReCS spectrum has  flux levels within   25\% of those obtained within a 4 arcsec aperture by Roche et al., who detected a deep silicate absorption band with $\tau_{9.7\umu{\rm m}} \sim$4.3.  The 3~$\mu$m spectrum of NGC 3094 has been presented by Imanishi  (2000) who found a weak 3.4~$\umu$m hydrocarbon absorption band with a peak optical depth, $\tau_{3.4\umu{\rm m}} \sim$ 0.04. Imanishi also claimed to detect a weak 3.3~$\umu$m PAH emission band; we see no evidence for the corresponding PAH emission features in the 8-13~$\umu$m spectrum.

 \subsection{NGC  5506}

%Schmitt et al 2001 ApJS 132 199     According to Braatz, Wilson, \& Henkel (1996) this galaxy has an H2O maser. The radio emission shows a linear structure along the east-west direction, surrounded by diffuse emission, with a total extension of 300 pc. The VLBA image from Roy et al. (2000) shows a double radio source along P.A. = 70, which will be used by Kinney et al. (2000). This galaxy is almost edge-on, and the H I image (Gallimore et al. 1999) shows absorption against the nuclear source. The hard X-ray spectrum has a flux F2-10 keV = 8.38 ? 10-11 ergs cm-2 s-1 and is absorbed by a column density NH = 3.4 ? 1022 cm-2 (Bassani et al. 1999). Maiolino et al. (1994) detected double-peaked line profiles, which they interpreted as outflow along P.A. = -16. Colbert et al. (1998) detected extended soft X-ray emission along the same direction, P.A. = -20, which is also coincident with the extended radio emission.

 NGC 5506 is a Sa spiral galaxy hosting an intermediate  Seyfert 1.9 nucleus at a redshift of 0.0062. The galaxy has a prominent dust lane and is almost edge-on, with an inclination angle approaching 90$^\circ$.   The nucleus is very compact in the mid-infrared,  and has an optical depth $\tau_{9.7\umu{\rm m}} \sim$ 1.4 and a prominent [SIV] emission line at 10.5~$\umu$m (Roche et al 1991, Siebenmorgen et al 2004).  The nucleus is classed as a Seyfert 1.9 as broad wings are detected on the infrared lines (Blanco, Ward \& Wright 1990, Nagar et al 2002).  The X-ray spectrum suggests substantial obscuration, with a  column of material corresponding to $\sim 3.4 \times 10^{22} {\rm cm}^{-2}$ (Risaliti et al 2000).  The hydocarbon absorption band at 3.4~$\umu$m is weak with an optical depth $\tau_{3.4\umu{\rm m}} \sim$ 0.04 (Imanishi 2000).  The TReCS acquisition image shows a bright compact core but the spectrum obtained at PA 0$^\circ$ has a flux level about 50\% lower  than detected by Roche et al (1991) in a 4 arcsec aperture, consistent with a compact core within a more extended emitting region.  The [SIV] line emission is more extended than the dust continuum emission along PA 0$^\circ$, corresponding approximately to the ionization cone axis found by Wilson, Baldwin \& Ulvestad (1985). 
 
 \subsection{NGC 7172}
 
 %Morganti et al 1999 A&AS..137..457:  
 %NGC 7172: This object has a linear radio structure (Figs. 3 and 4) whichis elongated roughly east-west (PA_radio_ = 90^deg^). A 13 cm core flux of 3 mJy has been detected with the PTI (R94).
 
 NGC 7172 is a type 2 Seyfert in a nearly edge-on Sa spiral galaxy at a redshift of 0.0087 (Sharples et al 1984).  It displays a prominent dust lane    running almost east-west and the Spitzer IRAC image shows mid-infrared emission extending over several arcseconds in the same direction; there is a suggestion of extension along this same axis in the T-ReCS acquisition image.  The nucleus of NGC 7172 is heavily obscured in the X-ray region with N(H) $\sim 8.6 \times 10^{22} {\rm cm}^{-2}$, and in the mid-infrared with a silicate absorption band with $\tau_{9.7\umu{\rm m}} \sim$ 2.8 (Roche et al 1991).  The hydrocarbon absorption band at 3.4~$\umu$ has an optical depth $\tau_{3.4\umu{\rm m}} \sim$ 0.09 (Imanishi 2000).  The T-ReCS image indicates that the core is extended on a scale of $\sim$1 arcsec at a position angle of 90$^\circ$.   The flux captured by the T-ReCS slit at a PA of 60$^\circ$ is about one third that detected by Roche et al in their 4 arscec aperture, consistent with a compact core within the more extended emitting source detected by Spitzer. 
 
\begin{table*}

%\begin{center}
Table 1.  Absorption features in nearby galaxies

\hspace*{3 mm}

\begin{tabular} {l c c l l c}
		 & $\tau_{9.7\umu{\rm m}} \mu$ Cep& $\tau_{9.7\umu{\rm m}}$ Trap & $\chi^2$/N ($\mu$ Cep)&  $\chi^2$/N (Trap) & $\tau_{3.4\umu{\rm m}}$$^a$  \\[2 mm]
NGC 4418$^b$  &   7.5$\pm$0.2 	& 6.6$\pm$0.2  &  2.8 & 6.4 \\ 
NGC 3094 &    4.7$\pm$0.1 	& 5.0$\pm$0.1   & 4.1 & 9.5 & 0.05 \\
NGC 7479$^c$  &   3.2$\pm$0.2 & 3.3$\pm$0.2 & 1.2 & 1.4 & 0.18 \\
NGC 7172  &   3.2$\pm$0.1     &  3.2$\pm$0.1& 3.9 &  6.9  &  0.08 \\
Circinus$^a$      &   1.76$\pm$0.1	& 2.4$\pm$0.1 & 4.63   & 4.2 \\
NGC 5506  &   1.1$\pm$0.05 	& 1.4$\pm$0.05 & 2.8 &  1.4 & 0.02: \\
NGC 1068$^a$  &   0.51$\pm$0.02 & 0.69$\pm$0.02  & 1.54   & 1.31 & 0.09 \\[2mm]
NGC 5506 0.6" N &   1.00$\pm$0.05 	& 1.27$\pm$0.05 & 2.3 &  2.2 &  \\
NGC 5506 0.3" N &   1..08$\pm$0.04 	& 1.40$\pm$0.04 & 2.7 &  1.8 &  \\
NGC 5506 0.3" S  &   1.33$\pm$0.04 	& 1.67$\pm$0.04 & 2.5 &  1.6 &  \\
NGC 5506 0.6" S &   1.31$\pm$0.05 	& 1.70$\pm$0.05 & 2.6 &  2.2 &  \\[3mm]
\end{tabular}

%\end{center}

\noindent 
{Notes:  
$^a$ $\tau_{3.4\umu{\rm m}}$ taken from Mason et al 2004.  
Silicate absorption depths taken from  \\ 
$^b$ Roche et al 1986;  $^c$ Roche et al 1991; $^d$ Roche et al 2006; $^e$ Roche et al 1984.}

\end{table*}

\subsection{Fits to the spectra}

In order to quantify the behaviour of the mid-IR absorption, we have fit the spectra with different emission and absorption components.  The fits cannot be physically realistic in this complex region, but do allow us to extract some quantitative data. 

Following Aitken \& Roche (1982) we have fit the spectra with emission spectra suffering extinction by cool silicate grains.  In these galaxies, inclusion of a PAH emission component does not improve the fits as indicated by the $\chi^2$ values and so the results presented employ emission and absorption spectra consisting of black-body and silicate grains.  We used silicate grain profiles derived from the Trapezium in Orion and the M supergiant $\mu$ Cephei, which are taken to be representative of molecular clouds  and the diffuse Galactic interstellar medium respectively (Roche \& Aitken 1984, Whittet et al 1988).    The goodness of fit is estimated from reduced $\chi^2$ values, with the errors estimated from the scatter in the data points. These allow us to differentiate between different combinations, but the range of values giving  adequate fits can be quite large for some  spectra.   In particular, the depth of the silicate absorption depends critically on the assumed spectral properties of the underlying emission. Here we assume that the underlying emission can be represented by grains with a blackbody emission spectrum.
  
For the most heavily obscured objects, NGC 3094 and NGC 7172, the reduced $\chi^2$ values indicate that a $\mu$ Cep-like emissivity provides a much better fit than a Trapezium-like emissivity profile. The results are  listed in Table 2 and inspection of Fig 2 indicates that the $\mu$ Cep curve provides a qualitatively much better fit to the curvature in the 8-9$\mu$m region than the Trapezium curve.  This is in line with previous conclusions for deeply-embedded objects, e.g. for NGC 4418 (Roche et al 1986) or IRAS 08572+3915 (Dudley \& Wynn-Williams 1997) and from Spitzer results (Spoon et al 2006).   The dust towards the deeply embedded galaxies is well-represented by dust very similar to that in the diffuse ISM in the Galaxy (Roche \& Aitken 1985, Chiar \& Tielens 2006).  However, the fit to NGC 3094 shows additional structure near 11$\mu$m which may represent a contribution from crystalline grains.  Spoon et al (2006) have detected absorption features at 11, 16, 19, 23 and 28$\mu$m which they attribute to crystalline silicates in the Spitzer IRS spectra of  galaxies with deep silicate absorption bands with $\tau_{9.7\umu{\rm m}} >$ 2.9.  Unfortunately, the S/N ratio of the present data are insufficient to determine whether the additional 11$\mu$m crystalline absorption is distributed across the emitting region or whether it is centrally concentrated.

For NGC 5506,  the silicate absorption feature is much shallower with $\tau_{9.7\umu{\rm m}} \sim$1.4 with the Trapezium profile or 1.1 with the  $\mu$ Cep profile towards the nucleus. As found for several other galaxies with moderate silicate absorption (e.g. in the Circinus galaxy; Roche et al 2006), the Trapezium curve provides the better fit.

\section{Discussion}

\subsection{Spectral variations in NGC 5506}
The dust continuum emission in NGC 5506 is compact but sufficiently bright to allow us to measure spectral differences  in the line emission and silicate absorption depth over the central 1.2 arcsec. At the distance implied by the redshift of NGC 5506, this corresponds to a total distance of about 170pc.  
The spectra to the north have shallower silicate absorption than on the nucleus with  $\tau_{9.7\umu{\rm m}} \sim$ 1.25 at 0.6 acsec North, while the optical depth is higher at  $\tau_{9.7\umu{\rm m}} \sim$ 1.7 at 0.6 arcsec south using the Trapezium profile; the colour temperature of the underlying emission is $\sim$300 K.  The [SIV] emission line intensity decreases by a factor 3 at 0.6 arcsec north and by a factor 10 at 0.6 arcsec south compared to the central position, reflecting the changes in silicate absorption depth and the corresponding extinction.  The continuum emission decreases more quickly than the line emission with distance from the nucleus so that the equivalent width of the [SIV] line increases.   The [SIV] intensity is lower on the south side of the nucleus, consistent with greater extinction than to the north if the intrinsic distribution is symmetric.   The total [SIV] flux detected in the TReCS slit, 1.7 10$^{-20}$ W cm$^{-2}$,  is 30 percent of that detected by ISO in a large ($\sim$ 14 x 20 arcsec) aperture (Sturm et al 2002) and about half that detected by Siebenmorgen et al (2004) in a 3 arcsec slit.  The 12.8~$\umu$m [Ne II] line is not detected in any of the TReCS spectra, suggesting that the [NeII] emission line detected by ISO  arises from more spatially extended, lower ionization gas than the [SIV] emission.

These spatial trends are consistent with an inclined obscuring layer that increases the extinction to both dust and line emitting regions to the south of the nucleus and results in lower extinction to the north, and which extends over at least 80~pc.   The variations in the depth of the silicate absorption band and the [SIV] emission line are qualitatively similar to those found in NGC 1068 by Mason et al (2006) and Rhee \& Larkin (2006)  and in Circinus by Roche et al (2006) who also found variations in the silicate absorption depth on subarcsec scales and line emitting regions more extended than the continuum emission.    Adopting an A(V) to $\tau_{9.7\umu{\rm m}}$ ratio of 18, as found in the diffuse ISM in the Galaxy (Roche \& Aitken 1984, Chiar \& Tielens 2006) indicates that the corresponding visual extinctions at 0.6 acsec S, on the nucleus and 0.6 arcsec N are A$_V \sim$ 30 mag, 25 mag and 22 mag respectively.

\subsection {Absorbing Columns}

The fits to the galaxy spectra hold only for simple models of a warm underlying emitting nucleus suffering extinction by  cold layers. Nonetheless, the match between the silicate absorption profile between $8-9\umu$m and $11.5-13\umu$m in NGC 3094 and that of $\mu$ Cep, representing the profile in the diffuse ISM in the Galaxy, is remarkable.  It is noteworthy  that the galaxies with very deep silicate absorption bands have profiles similar to the diffuse ISM while those with shallower silicate absorption are better matched by the molecular cloud  (Trapezium) curve. This is illustrated in Table 1, where the 9.7~$\umu$m absorption depths are listed together with the $\chi^2$ values obtained with the different silicate profiles for the galaxies discussed here, along with other examples of nearby AGN from the literature.  It seems that obscuring columns of $\tau_{9.7\umu{\rm m}}\ge$ 3 are better fit by the $\mu$ Cep curve, suggesting that molecular cloud  material does not produce most of the absorption. The origin of the difference between the Trapezium and $\mu$ Cep profiles is not clear  but it may result from mantles on grains or, more likely, grain size effects (e.g. see Bowey et al 1998).  We expect AGN to preferentially destroy small grains (Aitken \& Roche 1985, Voit 1992), and so we might expect the grains with Trapezium-like emissivities to be found closest to the AGN, and small grain destruction to be most complete in objects with the lowest absorbing columns.   Alternatively, it is possible that the silicate grains closest to the nuclei are located in very dense clumps, where they are cold and can accrete molecular material and may coagulate, and perhaps these regions are seen preferentially in the less obscured nuclei.

 NGC 5506 and NGC 7172 are close to edge-on to the line of sight, with prominent dust lanes in optical images, so it is perhaps likely that some of this extinction arises in  the galactic ISM.   The columns of material implied from the X-ray absorption ($\sim 3.4 \times 10^{22} $ and $\sim 8.6 \times 10^{22}  {\rm cm}^{-2}$ ; Risaliti et al 2000) are  similar to those inferred from the silicate absorption bands towards the nuclei  ($\sim 2.8 \times 10^{22}$ and $\sim 6.4 \times 10^{22}   {\rm cm}^{-2}$ for  NGC 5506 and NGC 7172 respectively,  estimated from the relationship between A(V) and $\tau_{9.7\umu{\rm m}}$ in the Galactic ISM (Roche \& Aitken 1984) and  a ratio of N$_H$ / A$_{V}$ = 1.9 10$^{21}$ cm$^{-2}$ mag$^{-1}$  (Bohlin, Savage \& Drake 1978).  The similar absorbing columns suggest  that the mid-IR and X-ray emitting  regions suffer similar extinctions in these objects, unlike many other AGN where the X-ray column is much higher than estimates from the IR.  Roche \& Aitken (1985) have found that the ratio of silicate absorption depth to visual extinction is higher towards the Galactic centre than towards more local objects in the Milky way, so the comparisons in absorbing column derived from different wavebands are uncertain by at least a factor 2. 
 
The 3.4$\mu$m absorption band is attributed to aliphatic hydrocarbons  and is detected along sightlines through the diffuse interstellar medium rather than through molecular material (e.g. Chiar et al 2000). 
The ratio of $\tau_{3.4\umu{\rm m}}$ to $\tau_{9.7\umu{\rm m}}$ provides estimates of the relative columns of hydrocarbons to silicates.  Along the line of sight  to the Galactic centre and more local sightlines within the Galaxy  this ratio is $\sim$ 0.06 and does not show significant variations (Chiar et al 2000).  The ratio in the galaxies listed in table 1  varies between 0.01 and 0.17.  If the 3.4$\mu$m band is representative of diffuse ISM material, we might expect the ratio of $\tau_{3.4\umu{\rm m}}$ to $\tau_{9.7\umu{\rm m}}$ to be closer to the Galactic value  in the galaxies that show the narrow $\mu$~Cep like silicate profile, but this is not the case.  In the most heavily obscured  galaxies, the 3.4~$\mu$m hydrocarbon absorption band is weaker relative to the silicate absorption band than in our Galaxy (see also Imanishi 2000). This may reflect differences in grain populations, perhaps as a result of irradiation by the AGN.  Spectropolarimetry shows that the grains producing the 3.4~$\mu$m absorption towards the Galactic centre are unpolarized in contrast to those producing the silicate absorption (Chiar et al 2006, Aitken et al 1986), indicating that the hydrocarbon and silicate absorption bands arise from separate populations.    However, contributions from different components, differences in the emitting volumes at 3 and 10~$\mu$m and complex circumnuclear structures may also be important.

\section{Conclusions}

NGC 5506 shows variations in silicate absorption depth on a scales of tens of  parsecs.  This could arise in a dusty structure extending over many tens of parsecs in the nuclear region.  High spatial resolution observations of NGC 1068 and Circinus show similar effects,  so that such structures  appear to be a common property of circumnuclear material in nearby AGN. 

The 10~$\mu$m silicate absorption  in galaxies with deep absorption closely resembles the  silicate profile in the diffuse interstellar medium in our Galaxy, as represented by the circumstellar emission from the red supergiant $\mu$~ Cep while less heavily obscured nuclei appear to be better fit by the Trapezium emissivity curve, which is representative of silicates in molecular clouds.  This could reflect preferential destruction of small grains in the galaxies with the lowest absorbing columns.  However, the absorption depth of the 3.4$\mu$m aliphatic hydrocarbon band, which is prominent in the Galactic diffuse ISM is not well correlated with the silicate absorption depth, even in galaxies with diffuse-ISM-like  profiles.  

In common with many heavily obscured ULIRGS,  NGC 3094 shows evidence of spectral structure at 11~$\mu$m which may arise from a population of crystalline silicates; higher quality spectra may be able to determine whether these grains are confined to the nuclear region  or are more widely distributed.

\section*{Acknowledgments}

This work is based on observations obtained at the Gemini Observatory, which is operated by the
Association of Universities for Research in Astronomy, Inc., under a cooperative agreement
with the NSF on behalf of the Gemini partnership: the National Science Foundation (United
States), the Particle Physics and Astronomy Research Council (United Kingdom), the
National Research Council (Canada), CONICYT (Chile), the Australian Research Council
(Australia), CNPq (Brazil) and CONICET (Argentina).  CP acknowledges support from 
NSF grant 0206617.

We thank the staff of Gemini-S, and especially Tom Hayward, for their assistance in collecting these data.

\label{lastpage}


\begin{thebibliography}{29}

\bibitem[\protect\citeauthoryear{Aitken}{1986}]{b1}Aitken, D. K. Briggs, G. P.  Roche, P. F.  Bailey, J. A. Hough, J. H., 1986, MNRAS, 218, 363
\bibitem[\protect\citeauthoryear{Aitken}{1982}]{b2}Aitken D.K.,  Roche P.F., 1982, MNRAS 200, 217
\bibitem[\protect\citeauthoryear{Aitken}{1985}]{b3}Aitken D.K.,  Roche P.F., 1985, MNRAS 213, 777
\bibitem[\protect\citeauthoryear{Antonucci}{1993}]{b4} Antonucci, R., 1993. ARAA 31, 473
\bibitem[\protect\citeauthoryear{Blanco}{1990}]{b5}Blanco, P.R., Ward  M.J., Wright G.S., 1990, MNRAS 242, 4P
\bibitem[\protect\citeauthoryear{Bohlin}{1978}]{b6}Bohlin, R.C., Savage, B.D., Drake J.F. 1978, ApJ, 224, 132
\bibitem[\protect\citeauthoryear{Bowey}{1998}]{b7}Bowey, J.E., Adamson, A.J.,  Whittet, D.C.B. 1998, MNRAS, 298, 131
\bibitem[\protect\citeauthoryear{Casassus}{2000}]{b8}Casassus S., Roche P.F., Barlow M.J., 2000, MNRAS, 314, 657.
\bibitem[\protect\citeauthoryear{Chiar}{2000}]{b9} Chiar, J. E. Tielens, A. G. G. M., Whittet, D. C. B., Schutte, W. A.  Boogert, A. C. A.  Lutz, D.  van Dishoeck, E. F.  Bernstein, M. P., 2000, ApJ, 537, 749
\bibitem[\protect\citeauthoryear{Chiar}{2006}]{b10}Chiar J.E., Tielens A.G.G.M., 2006, ApJ, 637, 774
\bibitem[\protect\citeauthoryear{Chiar}{2006b}]{b11}Chiar J.E., Adamson A.J., Whittet D.C.B., Chrysostomou A., Kerr T.H., Mason R.E., Roche P.F.,  Wright G., 2006, ApJ, 651, 268
\bibitem[\protect\citeauthoryear{Cohen}{1996}]{b12}Cohen M., Witteborn F.C., Bregman J.D., Wooden D.H., Salama A., Metcalfe L., 1996, AJ 112, 241.
\bibitem[\protect\citeauthoryear{Condon}{1990}]{b13}Condon, J. J., Helou G., Sanders D. B., Soifer, B. T. 1990, ApJS, 73, 359
\bibitem[\protect\citeauthoryear{Dudley}{1997}]{b14} Dudley C.C., Wynn-Williams C.G. 1997.  ApJ 488, 720.
\bibitem[\protect\citeauthoryear{Imanishi}{2000}]{b15}Imanishi, M. 2000, MNRAS, 319, 331
\bibitem[\protect\citeauthoryear{Imanishi}{2003}]{b16}Imanishi, M.2003, ApJ, 599, 918
\bibitem[\protect\citeauthoryear{Imanishi}{2006}]{b17} Imanishi M., Dudley C. C., Maloney P.R., 2006, ApJ, 637, 114
\bibitem[\protect\citeauthoryear{Mason}{2004}]{b18} Mason R. E., Wright G., Pendleton Y., Adamson A.,
2004, ApJ, 613, 770
\bibitem[\protect\citeauthoryear{Mason}{2006}]{b19} Mason R. E.,  Geballe T. R.,  Packham C.,  Levenson N. A.,  Elitzur M., Fisher R. S.,  Perlman E. 2006, ApJ, 640, 612
\bibitem[\protect\citeauthoryear{Nagar}{2002}]{b20} Nagar N. M., Oliva E.,  Marconi A.,  Maiolino R., 2002, A\&A, 391, 21
\bibitem[\protect\citeauthoryear{Rhee}{2006}]{b21} Rhee J.H., Larkin J.E., 2006, ApJ, 640, 625
\bibitem[\protect\citeauthoryear{Risaliti}{1999}]{b22} Risaliti G., Maiolino R.,  Salvati M, 1999, ApJ, 522, 157
\bibitem[\protect\citeauthoryear{Roche}{1984}]{b23} Roche P.F., Aitken D.K.,   1984, MNRAS, 208, 481
\bibitem[\protect\citeauthoryear{Roche}{1985}]{b24} Roche P.F., Aitken D.K.,   1985, MNRAS, 215, 425
\bibitem[\protect\citeauthoryear{Roche et al}{1986}]{25} Roche P. F.,  Aitken D. K., Smith C. H., James S.D. 1986, MNRAS, 218, 19P
\bibitem[\protect\citeauthoryear{Roche et al.}{1991}]{b26} Roche P.F., Aitken D.K., Smith C.H., Ward M.J.,  1991, MNRAS, 248, 606
\bibitem[\protect\citeauthoryear{Roche et al}{1984}]{27} Roche P. F.,  Aitken D. K., Whitmore, B.,  Phillips,  M., 1984, MNRAS, 207, 35
\bibitem[\protect\citeauthoryear{Roche et al.}{2006}]{b28} Roche P.F. et al., 2006, MNRAS, 367,1689
\bibitem[\protect\citeauthoryear{Siebenmorgen}{2004}]{b29} Siebenmorgen R., Krugel E., Spoon H.W.W., 2004, A\&A, 419, 49
\bibitem[\protect\citeauthoryear{Sharples}{1984}]{b30} Sharples R. M.,  Longmore, A. J., Hawarden, T. G., Carter, D, 1984, MNRAS, 209, 373
\bibitem[\protect\citeauthoryear{Spoon}{2006}]{b31} Spoon, H.W.W. et al. 2006, ApJ 638, 759
\bibitem[\protect\citeauthoryear{Sturm}{2002}]{b32} Sturm  E., Lutz, D., Verma, A., Netzer, H., Sternberg, A. Moorwood A. F. M., Oliva E., Genzel R. 2002. A\&A, 393,821
\bibitem[\protect\citeauthoryear{Telesco}{1998}]{b33}Telesco C.M., Pina R. K., Hanna K. T., Julian J. A., Hon D. B., Kisko T. M., 1998. SPIE 3354, 534
\bibitem[\protect\citeauthoryear{Voit}{1992}]{b34}Voit, G.M., 1992.  MNRAS, 258, 841
\bibitem[\protect\citeauthoryear{Whittet}{1988}]{b35}Whittet, D.C.B., Bode, M. F., Longmore, A. J., Adamson, A. J., McFadzean, A. D., Aitken, D. K., Roche, P. F., 1988, MNRAS, 233, 321
\bibitem[\protect\citeauthoryear{Wilson}{1985}]{b36} Wilson A.S., Baldwin, J.A., Ulvestad J.S., 1985, ApJ, 291. 627
\bibitem[\protect\citeauthoryear{Zenner}{1993}]{b37} Zenner S., Lenzen R., 1993. A\&A, 101, 363
\end{thebibliography}
\end{document}